\documentclass{llncs}
\usepackage{mathtools}
\usepackage{amssymb}
\usepackage{amsmath}
\usepackage{subfigure}
\usepackage{float}
\usepackage{xcolor}
\usepackage{hyperref}
\usepackage{stmaryrd}
\usepackage{algorithm}
\usepackage{algpseudocode}
\usepackage{wrapfig}
\usepackage{lineno}

\newcommand{\R}{\mathbb{R}}

\addtocounter{page}{0}

\renewcommand{\vec}[1]{\mathbf{#1}}

\renewcommand{\phi}{\varphi}

\renewcommand{\rm}{\rrbracket}

\newcommand{\ev}[1]{\diamondsuit_{#1}}
\newcommand{\glob}[1]{\boxempty_{#1}}

\newcommand{\kerPerc}[4]{\langle #1_{#2},#3 \rangle\, #4}
\newcommand{\kerSML}[4]{\langle #1_{#2},#3 \rangle\, #4}

\newcommand{\eSTL}{\text{SCL}}
\newcommand{\eSTLname}{\text{Signal Convolution Logic}}

\begin{document}
	\title{\eSTLname}
	\titlerunning{Signal Convolution Logic}  
	%
	\author{Simone Silvetti\inst{1,2} \and Laura Nenzi\inst{3} \and 
            Ezio Bartocci\inst{3} \and Luca Bortolussi\inst{4} }
	\authorrunning{S. Silvetti} 
	%
	\tocauthor{Simone Silvetti}
	\institute{
		DIMA, University of Udine, Udine, Italy
		\and
		Esteco S.p.A., Italy\\
		\email{simone.silvetti@gmail.com}
        \and 
        TU Wien, Vienna, Austria\\
        \email{\{laura.nenzi,ezio.bartocci\}@tuwien.ac.at}
        \and
        DMG, University of Trieste, Trieste, Italy\\
        \email{luca@dmi.units.it}
	}
	
	\maketitle
\begin{abstract}
We introduce a new logic called {\it \eSTLname} ($\eSTL$) that combines temporal 
logic with convolutional filters from digital signal processing.  $\eSTL$ enables
 to reason about the percentage of time a formula is satisfied in a bounded interval.  
We demonstrate that this new logic is a suitable formalism to effectively express non-functional 
requirements in Cyber-Physical Systems displaying noisy and irregular behaviours.
We define both a qualitative and quantitative semantics for it, providing an efficient 
monitoring procedure. Finally, we prove $\eSTL$  at work to monitor the {\it artificial pancreas} 
controllers that are employed to automate the delivery of insulin for patients with type-1 diabetes.


\end{abstract}

{\let\thefootnote\relax\footnotetext{
\scriptsize {E.B.\ and L.N.\ acknowledge the partial support of the Austrian National 
Research Network  S 11405-N23 (RiSE/SHiNE) of the Austrian Science 
Fund (FWF). E.B.,\ L.N.\ and S.S.\ acknowledge the partial support of the ICT COST Action IC1402 (ARVI).}}}

\section{Introduction}
\label{sec:intro}

Cyber-Physical Systems (CPS) are engineering, physical and 
biological systems tightly integrated with networked computational 
embedded systems monitoring and controlling the physical substratum.
The behaviour of CPS is generally modelled as a hybrid system
where the flow of continuous variables (representing the state of the physical components) 
is interleaved with the occurrence of discrete events (representing 
the switching from one mode to another, where each mode may model a different continuous dynamics).  
The noise generated by sensors measuring the data plays an important role in the 
modes switching and it can be captured using a stochastic extension of hybrid systems.

The exhaustive verification for these systems is in general undecidable. 
The available tools for reachability analysis are based on over-approximation of 
the possible trajectories and the final reachable set of states may result 
too coarse (especially for nonlinear dynamics) to be meaningful.
A more practical approach is to simulate the system and to monitor 
 both the evolution of the continuous and discrete state variables 
with respect to a formal requirement that specifies  
the expected temporal behaviour (see~\cite{chapter5} for a comprehensive survey).

Temporal logics such as Metric Interval Temporal Logic (MITL)~\cite{MalerN04} 
and its signal variant, Signal Temporal Logic (STL)~\cite{robust1}, are 
powerful formalisms suitable to specify in a concise 
way complex temporal properties.  In particular, STL enables to reason 
about real-time properties of components that exhibit both discrete and 
continuous dynamics.  The Boolean semantics of STL  decides 
whether a signal is correct or not w.r.t.  a given specification.
However, since a CPS model approximates the real system,  
the Boolean semantics is not always suitable to reason about its
behaviour, because it is not tolerant to approximation errors or to 
uncertainty. 

More recently, several notions of quantitative semantics (also called 
robustness)~\cite{robust1,fainekos-robust,filtering} have been 
introduced to overcome this limitation. These semantics enrich the expressiveness 
of Boolean semantics, passing from a Boolean concept of satisfaction (yes/no) 
to a (continuous) degree of satisfaction.  This allows us to quantify 
``how much" (w.r.t. a given notion of distance) a specific 
trajectory of the simulated system satisfies a given requirement. 
A typical example is the notion of robustness introduced by 
Fainekos et al. in~\cite{fainekos-robust}, where the binary satisfaction 
relation is replaced with a quantitative robustness degree function. 
The positive or negative sign of the robustness value indicates 
whether the formula is respectively satisfied or violated.
This notion of quantitative semantics is typically exploited in the falsification 
analysis~\cite{staliroauto,SankaranarayananF2012hscc,chapter5,avstl} to 
systematically generate counterexamples by searching, for example,
the sequence of inputs that would minimise the robustness towards the 
violation of the requirement.  On the other hand, the maximisation of
the robustness can be employed to tune the parameters of the 
system~\cite{eziotcs,stl-ps,chapter5,breach} to obtain a better 
resilience.  A more thorough discussion on other  
quantitative semantics will be provided in Section~\ref{sec:related}.  

\begin{figure}[!t]
	\centering
	\includegraphics[width=6.3cm]{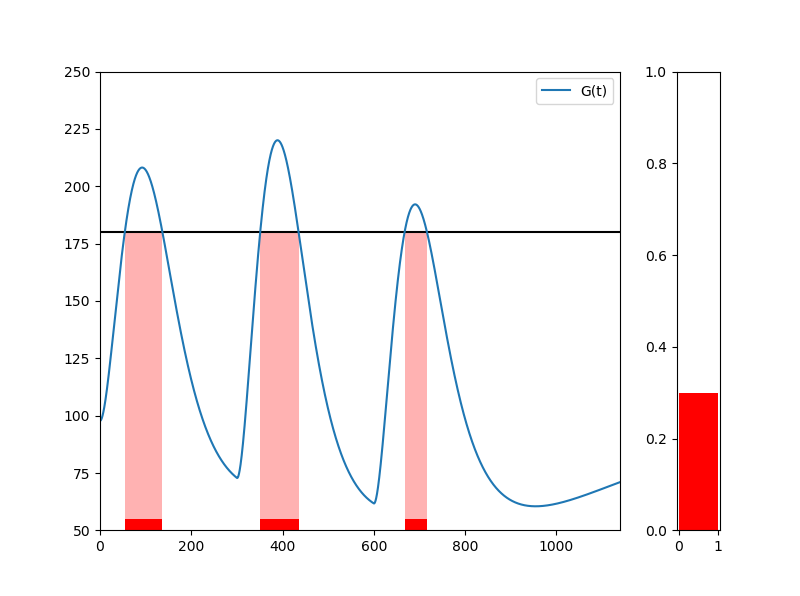}
	\hspace{-7mm}
	\includegraphics[width=6.3cm]{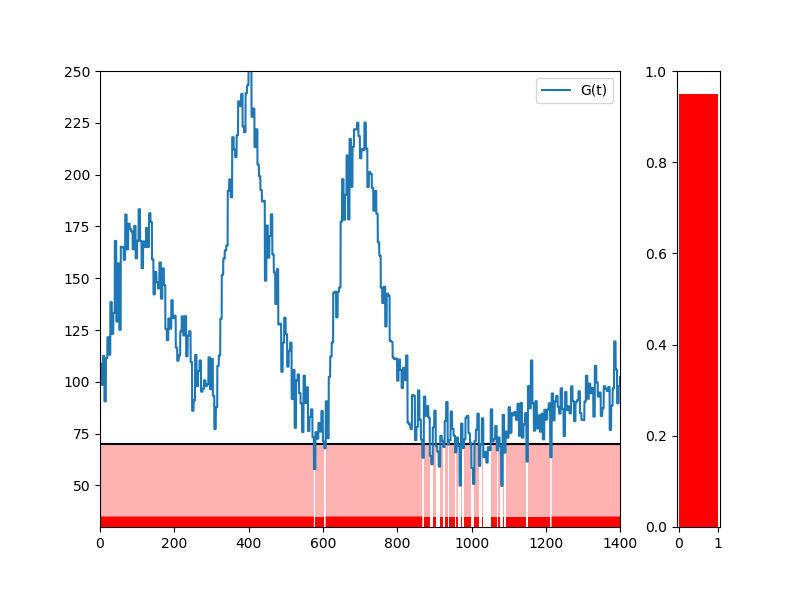}
    \vspace{-2ex}
	\caption{(\textbf{left}) A graphical representation of the property $\phi: \vec G(t)\geq180 $ for at least $12.5\%$ in [0,24h], meaning that the concentration of glucose has to be greater than 180 for at least 3h in 24h. (\textbf{right})  A graphical representation of the property $\psi: \vec G(t) > 70$ for at least 95\% in [0,24h]. The bars represents the percentage.
    }
	\label{fig:time}
\end{figure}

\paragraph{\bf Motivating Challenges.}
Despite STL is a powerful specification language, it does not come without limitations.  
An important type of properties that STL cannot express are the non-functional 
requirements related to the percentage of time certain events happen. 
The globally and eventually operators of STL can only check if a condition is true for all 
time instants or in at least one time instant, respectively. There are many real 
situations where these conditions are too strict, where it could be interesting 
to describe a property that is in the middle between eventually and always.
Consider for instance a medical CPS, e.g., a device measuring glucose level 
in the blood to release insulin in diabetic patients. In this scenario, we need to check if 
glucose level is above (or below) a given threshold for a certain amount of time, 
to detect critical settings.  Short periods under Hyperglycemia (high level of glucose) 
are not dangerous for the patient. An unhealthy scenario is when the patient remains 
under Hyperglycemia for more than 3 hours during the day, i.e., for $12.5\%$ of 24 
hours (see Fig. \ref{fig:time} left). This property cannot be specified by STL.
A second issue is that often such measurements are noisy, and  measurement 
errors or short random fluctuations due to environmental factors 
can easily violate (or induce the satisfaction) of a property.  One way to 
approach this problem is to filter the signal to reduce the impact of noise, 
This requires a signal pre-processing phase, which may however alter the signal 
introducing spurious behaviours. Another possibility, instead is to ask that the 
property is true for at least 95\% of operating time, rather than for 100\% of time, 
this requirements can be seen as a relaxed globally condition (see Fig. \ref{fig:time} right).
Finally, there are situations in which the relevance of events may change if they happen at different instants in a time window.
For instance, while measuring glucose level in blood, it is more dangerous if the 
glucose level is high just before meal, that means ``the risk becomes greater as 
we move away from the previous meal and approach the next meal". To capture this, 
one could give different weights if the formula is satisfied or not at the end or in the 
middle of a time interval, i.e., considering inhomogeneous temporal satisfaction 
of a formula. This is also not possible in STL.  
\vspace{-8mm}
%


\paragraph{\bf Contributions.}
In this paper, we  introduce a new logic based on a new temporal operator, $\langle k_T, p\rangle\phi$, 
that we call \emph{the convolution operator}, which overcomes these limitations.  It depends on 
a non-linear kernel function $k_T$, and requests that the convolution between the kernel and the 
signal (i.e., the satisfaction of $\phi$) is above a given threshold $p$.
This operator allows us to specify queries about the fraction of time a certain property is satisfied, 
possibly weighting unevenly the satisfaction in a given 
time interval $T$, e.g., allowing to distinguish traces that satisfy a property in specific parts of $T$. 
We provide a Boolean 
semantics, and then define  a quantitative semantics, proving its soundness 
and correctness with respect to the former. 
Similarly to STL, our definition of quantitative semantics permits to quantify the maximum allowed uniform translation of the signals preserving the true value of the formula. 
We also show that $\eSTL$  is strictly more expressive than $STL(\ev{},\glob{})$ (the fragment of STL which considers only eventually $\ev{}$ and globally $\glob{}$ operators) and 
then we provide the monitoring algorithms for both semantics.
Finally, we show  $\eSTL$  at work to monitor the behaviour of 
an artificial pancreas device releasing insulin in patients 
affected by type-I diabetes.

\paragraph{\bf Paper structure.}
The rest of the paper is organized as follows. In Section~\ref{sec:related} we 
discuss the related work. Section~\ref{sec:background} provides the necessary 
preliminaries.  Section~\ref{sec:scl}  presents the syntax and the semantics of 
$\eSTL$ and discuss its expressiveness. In Section~\ref{sec:monitoring}, 
we describe our monitoring algorithm and in Section~\ref{sec:casestudy} we show an 
application of $\eSTL$ for monitoring an insulin releasing device in diabetic patients. 
Finally, we draw final remarks in Section~\ref{sec:conclusion}.

\section{Related Work}
\label{sec:related}

The first quantitative semantics,
introduced by Fainekos et al.~\cite{fainekos-robust} and then used by 
Donze et al.~\cite{robust1} for STL, is based on the notion of {\em spatial 
robustness}. Their approach  
replaces the binary satisfaction relation with a function returning a  real-value representing the distance from the unsatisfiability set in terms of the uniform norm. 
In~\cite{robust1} the authors consider 
 also the displacement of a  signal in the time domain (temporal robustness). 
These semantics, since are related with the
uniform-norm, are very sensitive to glitches (i.e., 
sporadic peaks in the signals due to measurement 
errors).

To overcome this limitation Rodionova et al.~\cite{filtering} 
proposed a quantitative semantics based on filtering. 
More specifically they provide a quantitative semantics for 
the \emph{positive normal form} fragment of STL
which measures the number of times a formula it 
is satisfied within an interval associating with different types of kernels.  
However, restricting the quantitative 
semantics to the positive normal form  
gives up the duality property between the 
eventually and the globally operators, and the correctness property, which instead are both kept in our approach. 
Furthermore, their work is just theoretical and there is no discussion on how to efficiently evaluate such a properties.

In~\cite{avstl},  Akazaki et al. have extended the syntax of STL by introducing averaged temporal operators. 
Their quantitative semantics expresses the preference that a specific requirement occurs as earlier as 
possible or for as long as possible, in a given time range. 
Such  time inhomogeneity can be evaluated only in the quantitative semantics (i.e. the new operators, at the Boolean level, are equal to the classic STL temporal operators).
Furthermore, the new operators force separations of two robustness (positive and negative) and it is lost also in this case the correctness property.

An alternative way to tackle the  noise of a signal is to consider explicitly their stochasticity. Recently, there has been a great effort to define several stochastic extensions of STL, such as Stochastic Signal Temporal 
Logic (StSTL)~\cite{Jiwei2017}, Probabilistic Signal Temporal Logic (PrSTL)~\cite{SadighK16} and
Chance Constrained Temporal Logic (C2TL)~\cite{JhaRSS18}. 
The type of quantification is intrinsically different,
while the probabilistic operators quantify on the signal values, our convolutional operator quantifies over the time in which the nested formula is satisfied.
Furthermore, all these approaches rely on the use of probabilistic atomic predicates that need to be quantified over the probability distribution of a model (usually a subset of samples). As such, they need computationally expensive procedures to be analyzed. Our logic, instead, operates directly on the single trace, 
without the need of any probabilistic operator, in this respect being closer to digital signal processing. 
%
\section{Background}
\label{sec:background}
In this section, we introduce the notions needed later in the paper: signals, kernels, and convolution.

\begin{definition}[Signal]
	\label{tab:kernel}
	A \emph{signal} $\vec s : \mathcal{T} \to \mathcal{S}$ is a function from an interval $\mathcal{T} \subseteq \mathbb{R}$ to a subset  $\mathcal{S}$ of $\mathbb{R}^n,\, n<+\infty$. Let us denote with $\mathcal{D}(\mathcal{T};\mathcal{S})$ a generic set of signals.
\end{definition}
When $\mathcal{S} = \{0,1\}$, we talk of Boolean signals. In this paper, we consider piecewise constant signals, represented by a sequence of time-stamps and values. Different interpolation schemes (e.g. piecewise linear signals) can be treated similarly as well.  


\begin{definition}[Bounded Kernel]
Let be $T \subset  \R$ a closed interval. We call bounded kernel a function $k_{T}\colon \R \to \R$ such that: 
\begin{equation}
\int_{T} k_T(\tau)d\tau=1 \quad  \text{and} \quad \forall t  \in T,\, k_{T}(t)>0.
\end{equation}
\end{definition}

\begin{table}[!t]
	\label{tab:kernel}
	\begin{center}
		\begin{tabular}{|c|c|}
			\hline
			\textbf{kernel} & \textbf{expression} \\
			\hline
			constant ($\mathtt{flat}(x)$) & $\mathtt{1}(x)/(T_1-T_0)$ \\
			\hline
				exponential ($\mathtt{exp}[\alpha](x)$)& $\exp(\alpha x )/\int_T \exp(\alpha \tau ) d\tau$  \\
				\hline		
			gaussian ($\mathtt{gauss}[{\mu,\sigma}] (x)$) &  $\exp((x-\mu)^2 )/\sigma^2)/\int_T \exp((x-\mu)^2 )/\sigma^2) d\tau$\\
			\hline
	
		\end{tabular}
	\end{center}
	\caption{Different kind of kernels.}
    \vspace{-3ex}
\end{table}
Several examples of  kernels are shown in Table~\ref{tab:kernel}. We call $T$ the time window of the bounded kernel $k_T$, which will be used as a convolution \footnote{This operation is in fact a cross-correlation,  but here we use the same convention of the deep learning community and call it convolution.} 
operator, defined as: 
\[ 
(k_T*f)(t) = \int_{t+T} k_T(\tau-t) f(\tau)d\tau
\]
We also write $k_T(t)*f(t)$ in place of $(k_T*f)(t)$. 

In the rest of the paper, we assume that the function $f$ is always a Boolean function: $ f\colon \R \to \{0,1\} $. This implies that $  \forall t \in \R,\, (k_T*f)(t) \in [0,1]$, i.e. the convolution kernel will assume a value in $[0,1]$ 
This value can be interpreted as a sort of \emph{measure} of how long the function $f$ is true in $t+T$. In fact, the kernel induces a measure on the time line, giving different importance of the time instants contained in its time window $T$. 
As an example, suppose we are interested in designing a system to make an output signal $f$ as true as possible in a time window $T$ (i.e., maximizing $k_T * f$). Using a non-constant kernel $k_T$ will put more effort in making $f$ true in the temporal regions of $T$ where the value of the kernel $k_T$ is higher. 
More formally, the analytical interpretation of the convolution is simply the expectation value of $f$ in a specific interval $t+T$ w.r.t. the measure $k_T(dx)$ induced by the kernel. In Fig.~\ref{fig:battle,quant} (a) we show some example of different convolution operators on the same signal.
\section{\eSTLname}
\label{sec:scl}
In this section, we present the syntax and semantics of  $\eSTL$, in particular of the new convolutional operator $\kerPerc{k}{T}{p}$, discussing also its soundness and correctness, and finally comment on the expressiveness of the logic.    
\paragraph{\bf Syntax and Semantics.}
The atomic predicates of $\eSTL$ are inequalities on a set of real-valued 
variables, i.e. of the form $\mu(\vec{s}) {:=} [g(\vec{s})\geq 0]$, where 
$g:\mathcal{S} \to \mathbb{R}$ is a continuous function, $\vec{s} \in \mathcal{S} $ 
and consequently  $\mu: \mathcal{S} \to \{\top,\bot\}$.
The well  formed formulas $\mathcal{L}_{\eSTL}$ of $\eSTL$ are defined by the following grammar:
 	\begin{equation}
 	\label{eq:grammarSTL}
 	\phi := \bot \,|\,\top \,|\, \mu \,|\, \neg \phi \,|\,  \phi \vee \phi \,|\,  \kerPerc{k}{T}{p} \phi,
 	\end{equation}
 where $\mu$ are atomic predicates as defined above, $k_T$ is a bounded kernel and $p\in[0,1]$. 
  $\eSTL$ introduces the novel convolutional operator $\langle k_T, p\rangle \phi$ (more precise, a family of them) 
defined parametrically w.r.t. a kernel $k_T$ and a threshold $p$.
This operator specifies  the probability of $\phi$ being 
true in $T$, computed w.r.t. the probability measure 
$k_T(ds)$ of $T$, the choice of different types of kernel $k$ will give rise to different kind of operators (e.g. a constant kernel will measure the fraction of time $\phi$ is true in $T$, while an exponentially decreasing kernel will concentrate the focus on the initial part of $T$). As usual, we interpret the $\eSTL$ formulas over signals.

Before describing the semantics, we give a couple of examples of properties.
Considering again the glucose scenario presented in Section~\ref{sec:intro}. The properties in Fig.~\ref{fig:time} are specified in $\eSTL$ as $\phi : \kerSML{\mathtt{flat}}{[0,24h]}{0.125}{\vec G(t) \ge 180}$, $\psi :  \kerSML{\mathtt{flat}}{[0,24h]}{0.95}{\vec G(t) \ge 70}$.
We can use instead an exponential increasing kernel to described the more dangerous situation of high glucose closed to the next meal, e.g. $\psi :  \kerSML{\mathtt{exp}}{[0,8h]}{0.95}{\vec G(t) \ge 180}$.

We introduce now the Boolean and quantitative semantics.  As the temporal operators $\langle k_T, p\rangle$ are time-bounded,  time-bounded signals are sufficient to assess the truth of every formula. 
 In the following, we denote with $\mathcal{T}(\phi)$ the minimal duration 
 of a signal allowing a formula $\phi$ to be always  evaluated. $\mathcal{T}(\phi)$ 
 is computed as customary by structural recursion.

\begin{definition}[Boolean Semantics]
 		\label{def:boolean_semantics}
 		Given a signal $\vec{s}\in \mathcal{D}(\mathcal{T};\mathcal{S})$, the Boolean semantics $\chi\colon \mathcal{D}(\mathcal{T};\mathcal{S}) \times \mathcal{T} \times \mathcal{L}_{\eSTL} \to \{0,1\}$ is defined recursively by:
\begin{subequations}
\begin{align}
\chi(\vec{s},t,\mu) &= 1  \iff \mu(\vec{s}(t))=\top \mbox{ where } \mu(X) \equiv [g(X) \ge 0]\\
\chi(\vec{s},t,\neg \phi) & = 1  \iff   \chi(\vec{s},t,\phi)=0 \label{op:neg} \\
\chi(\vec{s},t,\phi_1 \vee \phi_2 ) & = \max(\chi(\vec{s},t,\phi_1),\chi( \vec{s},t,\phi_2)) \\
\chi(\vec{s},t,\langle k_T, p\rangle \phi) &=1 \iff k_T(t) * \chi(\vec{s},t,\phi)\ge p
\label{op:int}
 		\end{align}
  		\end{subequations}
Moreover, we let $\chi( \vec{s}, \phi)=1 \iff \chi( \vec{s},0, \phi)=1$. 
\end{definition}


The atomic propositions $\mu$ are inequalities over the signal's variables. The semantics of negation and conjunction are the same as classical temporal logics. The semantics of $\langle k_T, p\rangle \phi$ requires to compute the convolution of $k_T$ with the truth value $\chi( \vec{s},t,\phi)$ of the formula $\phi$ as a function of time, seen as a Boolean signal, and compare it with the threshold $p$.

An example of the Boolean semantics can be found in Fig.~\ref{fig:battle,quant} ({\bf left - bottom}) where four horizontal bars visually represent the validity of  $\psi = \langle k_{[0, 0.5]}, 0.5\rangle (s >0)$, for 4 different kernels $k$ (one for each bar). We can see that the the only kernel for which $\chi( s, \psi)=1 $ is the exponential increasing one $k=\mathtt{exp}[3]$ .

{\sloppy 
\begin{definition}[Quantitative semantics]
	\label{def:quantiative_semantics}
	The quantitative semantics $\rho:\mathcal{D}(\mathcal{T};\mathcal{S}) \times \mathcal{T} \times \mathcal{L}_{\eSTL} \rightarrow \mathbb{R}$ is defined as follows:
	\begin{subequations}
	\begin{align}
		\rho(\vec{s},t,\top)&= +\infty\\
		\rho(\vec{s},t,\mu)&= g(\vec{s}(t))  \mbox{ where $g$ is such that } \mu(X) \equiv [g(X) \ge 0]\\
		\rho(\vec{s},t,\neg \phi)&= - \rho(\phi,\vec{s},t)\\
		\rho(\vec{s},t,\phi_1 \vee \phi_2)&=\max( \rho(\phi_1,\vec{s},t), \rho(\phi_2,\vec{s},t))\\
		\rho( \vec{s},t,\langle k_T,p\rangle \phi)&=\max \{r \in \R \mid k_T(t) * [\rho(\vec{s},t,\phi) > r]\ge p \}
		\label{def:qs:4}
	 \end{align} 
 \end{subequations}
 Moreover, we let $\rho(\vec{s},\varphi):=\rho(\vec{s},0,\varphi)$.
 \end{definition}
 }

where $[\rho(\vec{s},t,\phi) > r]$ is a function of $t$ such that $[\rho(\vec{s},t,\phi) > r] =1$ if $\rho(\vec{s},t,\phi) > r$, 0 otherwise.
Intuitively the quantitative semantics of a formula $\phi$ w.r.t. a primary signal $\vec s$ describes the maximum allowed uniform translation of the secondary signals $\vec{g}(\vec{s}) = (g_1(\vec{s}),\ldots,g_{n(\phi)}(\vec{s}))$ in $\phi$ preserving the truth value of $\phi$. Stated otherwise, a robustness of $r$ for $\phi$ means that all signals $\vec {s'}$ such that $\|\vec{g}(\vec{s'}) - \vec{g}(\vec{s})\|_{\infty}\leq r$ will result in the same truth value for $\phi$: $\chi(\vec{s},t,\phi) = \chi(\vec{s'},t,\phi)$. Fig.~\ref{fig:battle,quant}(b) shows this geometric concept visually. Let us consider the formula $\phi = \langle k_{[0,3]}, 0.3\rangle (\vec s >0)$, $k$ a flat kernel. A signal $s(t)$ satisfies the formula if it is greater than zero for at most the $30\%$ of the time interval $T=[0,3]$. The robustness value corresponds to how much we can translate $s(t)$ s.t. the formula is still true, i.e. $r$ s.t. $s(t)-r$ still satisfies $\phi$. In the figure, we can see that $r=0.535$. The formal justification of it is rooted in the correctness theorem (Theorem~\ref{th:correctness}).    

\vspace{-2mm}
\begin{figure}[h]
	\centering
\subfigure{\includegraphics[height=4.2cm]
{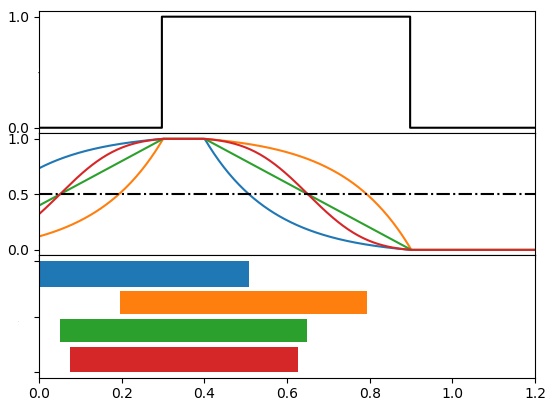}}
\subfigure{\includegraphics[height=4.2cm]
{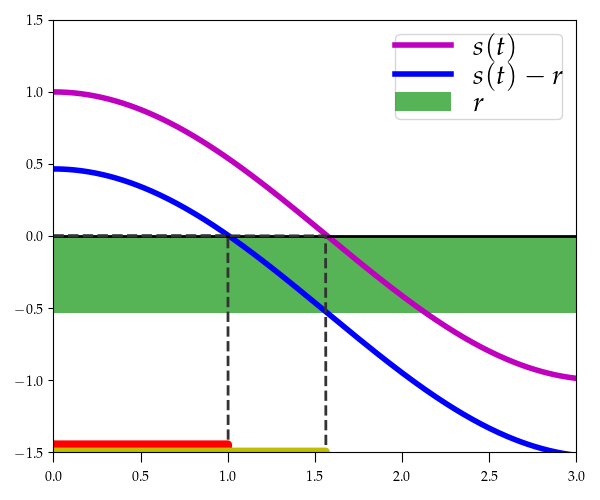}}
	\caption{(\textbf{left - top})  A Boolean signal $s(t)$ TRUE in $[0.3,0.9]$ and FALSE outside. (\textbf{left - middle}) Convolution of the kernel function 
    $({\mathtt{exp}[3]}_{[0,0.5]} * s)(t)$ (blue), 
    ${(\mathtt{exp}[-3]}_{[0,0.5]} * s)(t)$ (orange),
    $(\mathtt{flat}_{[0,0.5]} * s)(t)$ (green) and     $(\mathtt{gauss}_{[0,0.5]} * s)(t)$ (red) with the signal above in the time windows. The horizontal threshold is set to $0.5$. 
    (\textbf{left - bottom})  
    The 4 horizontal bars show when 
    $\chi( s, \psi,t)=1 $, with $\psi = \langle k_{[0, 0.05]}, 0.5\rangle (s >0)$, i.e when $(k_{[0,0.5]} * s)(t) > 0.5$.
    ({\bf right}) Example of quantitative semantics of $\eSTL$. A signal $s(t)$ satisfies the formula $\phi = \langle k_{[0,3]}, 0.3 \rangle (\vec s >0)$, with $k$ a flat kernel, if it is greater than zero for at most the $30\%$ of the time interval $T=[0,3]$. The robustness value corresponds to how much we can translate $s(t)$ s.t. the formula is still true, i.e. $\rho(s, \phi) = r$ s.t. $s(t)-r$ still satisfies $\phi$, (red line). In the figure we can see that $\rho(s, \phi)=0.535$.}
	\label{fig:battle,quant}
\end{figure} 

\paragraph{\bf Soundness and Correctness.}
\label{sec:soundcorrect}
We turn now to discuss soundness and correctness of the quantitative semantics with respect to the Boolean one. The proofs of the theorems can be found in the on-line version of the paper on arXiv.

\begin{theorem}[Soundness Property]
\label{th:soundness}
	The quantitative semantics is sound with respect to the Boolean semantics, than means:
	$$
		\rho(\vec{s},t, \phi) > 0 \implies (\vec{s},t) \models \phi 
       \qquad and \qquad 
		\rho(\vec{s},t, \phi) < 0 \implies (\vec{s},t) \not \models \phi
$$
\end{theorem}
\begin{definition}
	Consider a $\eSTL$ formula $\phi$ with atomic predicates $\mu_i:=[g_i(X)\ge 0 ],\, i\le n$, and signals $\vec s_1,\vec s_2 \in \mathcal{D}(\mathcal{T};\mathcal{S})$. We define
	 \[
	 \|\vec s_1-\vec s_2\|_\phi:=\underset{i\le n}{\max} \,\underset{t\in \mathcal{T}(\phi)}{\max}\, |g_i(\vec s_1(t))-g_i(\vec s_2(t)) |
	\]
\end{definition}

\begin{theorem}[Correctness Property] 
\label{th:correctness}
	The quantitative semantics $\rho$ satisfies the correctness property with respect to the Boolean semantics if and only if, for each formula $\phi$, it holds:
	\[  \forall \vec s_1,\vec s_2 \in \mathcal{D}(\mathcal{T};\mathcal{S}),\, \|\vec s_1-\vec s_2\|_\phi<\rho(\vec{s}_1,t,\phi)  \Rightarrow \chi(\vec s_1,t, \phi)=\chi( \vec s_2,t, \phi)
  \]	
\end{theorem}

\paragraph{\bf Expressiveness.}

We show that $\eSTL$ is more expressive than the fragment of STL composed of the logical connectivities and the eventually $\ev{}$ and globally $\glob{}$ temporal operators, i.e., $STL(\ev{},\glob{})$.

First of all, globally is easily definable in $\eSTL$. Take any kernel $k_T$, and observe that $\glob{T} \phi \equiv \langle k_T, 1\rangle \phi$, as $\langle k_T, 1\rangle \phi$ holds only if $\phi$ is true in the whole interval $T$. This holds provided that we restrict ourselves to Boolean signals of finite variation, as for~\cite{MalerN04}, which are changing truth value a finite amount of times and are never true or false in isolated points: in this way we do not have to care what happens in sets of zero measure. With a similar restriction in mind, we can define the eventually, provided we can check  that $k_T(t) * \chi( \vec{s},t,\phi) >0$.

\sloppy
To see how this is possible, start from the fundamental equation $k_T(t) * \chi( \vec{s},t,\neg \phi)=1-k_T(t) * \chi( \vec{s},t,\phi)$.
By applying \ref{op:int} and \ref{op:neg}  we  easily get
$ \chi(\vec{s},t,\neg \langle k_T, 1-p\rangle \neg \phi) = 1  \iff k_T(t) * \chi( \vec{s},t,\neg \phi) < 1-p
	\iff k_T(t) * \chi( \vec{s},t,\phi) > p$.
For compactness we write $\langle k_T, p\rangle^* = \neg \langle k_T, 1-p\rangle \neg $, and thus define the eventually modality as $\ev{T} \phi \equiv \langle k_T, 0\rangle ^*\phi$. By definition, this is the  dual operator of $\glob{T}$. Furthermore, consider the uniform kernel $\mathtt{flat}_T$: a property of the form  $\langle \mathtt{flat}_T, 0.5\rangle \phi$, requesting $\phi$ to hold at least half of the time interval $T$, cannot be expressed in STL, showing that $\eSTL$ is more expressive than  STL$(\ev{},\glob{})$. 

Note that defining a new quantitative semantics has an intrinsic limitation.  Even if the robustness can help the system design or the falsification process by guiding the underline optimization, it cannot be used at a syntactic level. 
It means that we cannot write logical formulas which predicate about the property.  For example, 
we cannot specify behaviors as \emph{the property has to be satisfied in at least the 50\% of interval I}, but we can only measure the percentage of time the properties has been verified.
Furthermore, lifting filtering and percentage at the syntactic level has other important two advantages. 
First, it preserves duality of eventually and globally operator, meaning that we are not forced to restrict our definition to  positive formulae, as in~\cite{filtering}, or to present two separate robustness measures as in~\cite{avstl}.
Second, it permits to introduce a quantitative semantics which  quantifies the robustness with respect to signal values instead of the percentage values and that satisfies the correctness property.


\section{Monitoring Algorithm}
\label{sec:monitoring}
In this section, we present the monitoring algorithms to 
evaluate the convolution operators $\langle k_T,p\rangle \phi$. For all the other operators we can rely on established algorithms as \cite{MalerN04} for Boolean monitoring and \cite{robust1} for the quantitative one. 

\paragraph{\bf Boolean Monitoring.}
We provide an efficient monitor algorithm for the Boolean 
semantics of $\eSTL$ formulas. 
Consider an $\eSTL$ formula $\langle k_{[T_0,T_1]}, p \rangle \phi$ and a signal $\vec s$. We are interested in computing $\chi ( \vec s , t,\langle k_{[T_0,T_1]},  p \rangle \phi) =[H(t) - p \ge 0]$, as a function of $t$, where $H$ is the following convolution function
\begin{equation}
\label{eq:monitoring}
H(t) = k_{T}(t) * \chi(\vec s ,t,\phi )=  \int_{t+T} k_{T} (\tau-t)  \chi(\vec s , \tau,\phi )  d\tau
\end{equation}

It follows that the efficient monitoring of the Boolean semantics of $\eSTL$ is linked to the efficient evaluation of $H(t)-p$, which is possible if  $H(t+\delta)$ can be computed by reusing the value of $H(t)$ previously stored.
To see how to proceed, assume the signal $\chi(\vec s, t,\phi )$ to be \emph{unitary}, namely that it is true in a single interval of time, say from time $u_0$ to time $u_1$, and false elsewhere. We remark that is always possible to decompose a signal in unitary signals, see \cite{MalerN04}.

\begin{figure}[!t]
\begin{center}
\includegraphics[width=0.8\textwidth]{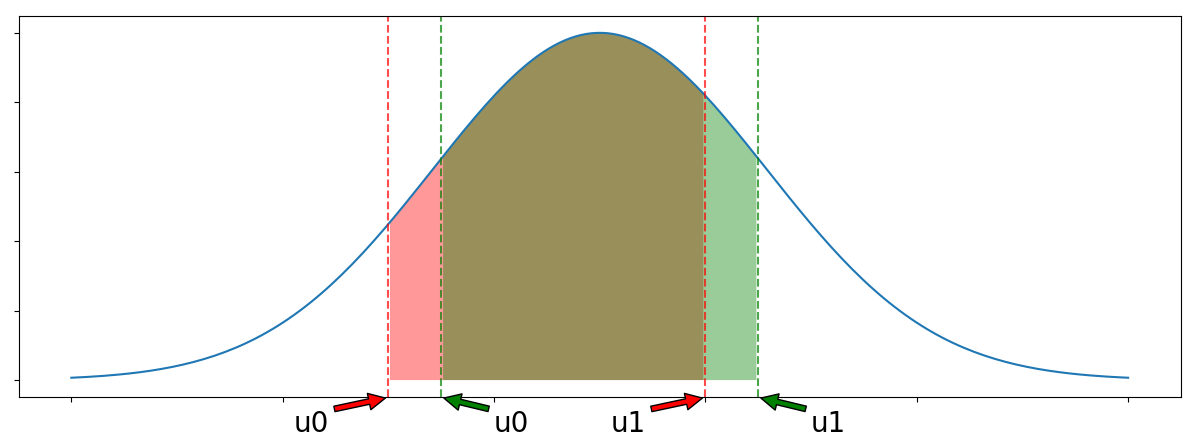}
\end{center}
\vspace{-0.5cm}
\caption{Sketch of the general monitoring algorithm. The green arrows represents $[u_0,u_1]$ in the constitutional window at time $t$, the red arrows instead represents the same interval at time $t+\delta$ (backwards translation).}
\label{fig:monitor}
\end{figure}

In this case, it easily follows that the convolution with the kernel will be non-zero only if the interval $[u_0,u_1]$ intersects the convolution window $t+T$. Inspecting Figure \ref{fig:monitor}, we can see that sliding the convolution window forward of a small time $\delta$ corresponds to sliding the positive interval of the signal $[u_0,u_1]$ of $\delta$ time units backwards with respect to the kernel window. In case $[u_0,u_1]$ is fully contained into $t+T$, by making $\delta$ infinitesimal and invoking the fundamental theorem of calculus, we can compute the derivative of $H(t)$ with respect to time as $\frac{d}{dt}H(t) = k_T(u_0 - t) - k_T(u_1 - t)$. By taking care of cases in which the overlap is only partial, we can derive a general formula for the derivative:
\begin{equation}
\label{eq:generalODEpositiveInterval}
\frac{d}{dt}H(t) = k_T(u_0 - (t+T_0))I\{u_0\in t+T\} - k_T(u_1 - (t+T_1))I\{u_1\in t+T\},
\end{equation}
where $I$ is the indicator function, i.e. $I\{u_i \in t+T\} = 1$ if $u_i \in t+T $ and zero otherwise.
This equation can be seen as a differential equation that can be integrated with respect to time by standard ODE solvers (taking care of discontinuities, e.g. by stopping and restarting the integration at boundary times when the signal changes truth value), returning the value of the convolution for each time $t$. The initial value is $H(0)$, that has to be computed integrating explicitly the kernel (or setting it to zero if $u_0\geq T_1$).
If the signal $\chi(\vec s , t,\phi )$ is not unitary, we have to add  a term like the right hand side of \refeq{eq:generalODEpositiveInterval} in the ODE of $H(t)$ for each unitary component (positive interval) in the signal. 
We use also a root finding algorithm integrated in the ODE solver to detect when the property will be true or false, i.e. when $H(t)$ will be above or below the threshold $p$.

The time-complexity of the algorithm for the convolution operator is proportional to the computational cost of numerically integrating the differential equation above. Using a solver with constant step size $\delta$, the complexity is proportional to the number of integration steps, times the number $N_U$ of unitary components in the input signal, i.e. $O(N_U(T_s/\delta))$. A more detailed description of the algorithm can be found in Appendix \ref{app:monitoring}.

\paragraph{\bf Quantitative Monitoring.}
In this paper, we follow a simple approach to monitor it: we run the Boolean monitor for different values of $r$ and $t$ in a grid, using a  coarse grid for $r$, and compute at each point of such grid the value $H(t,r) = k_T(t) * [\rho(\vec{s},t,\phi)  > r] - p$.   Relying on the fact that $H(t,r)$ is monotonically decreasing in $r$, we can find the correct value of $r$,  for each fixed $t$, by running a bisection search starting from the unique values $r_k$ and $r_{k+1}$ in the grid such that $H(t,r)$ changes sign, i.e. such that $H(t,r_k) < 0 < H(t,r_{k+1})$.  
The bounds of the $r$ grid are set depending on the bounds of the signal, and may be expanded (or contracted) during the computation if needed. Consider that the robustness can assumes only a finite number of values because of the finite values assumed by the pieacewise-constant inputs signals.
A more efficient procedure for quantitative monitoring is in the top list of our future work, and it can be obtained by exploring only a portion of such a grid, combining the method with the boolean monitor based on ODEs, and  alternating steps in which we advance  time from $t$ to $t+h$ (fixing $r_t$ to its exact value at time $t$), by integrating ODEs and computing $H(t+h,r_t)$, and steps in which we  adjust the  value of $r_t$ at time $t+h$ by locally increasing or decreasing its value (depending if $H(t+h,r_t)$ is negative or positive), finding $r_{t+h}$ such that  $H(t+h,r_{t+h})= 0$.

\section{Case Study: Artificial Pancreas }
\label{sec:casestudy}

In this example, we show how $\eSTL$ can be useful in the specification and monitoring of the Artificial Pancreas (AP) systems.  The AP is a closed-loop system of insulin-glucose for the treatment of Type-1 diabetes (T1D), which is a chronic disease caused by the inability of the pancreas to secrete insulin, an hormone essential to regulate the blood glucose level. In the AP system, a Continuous Glucose Monitor (CGM) detects the blood glucose levels and a pump delivers insulin through injection regulated by a software-based controller.

The efficient design of control systems to automate the delivery of insulin is still an open challenge  for many reasons. Many activities are still under control of the patient, e.g., increasing insulin delivery at meal times (meal bolus), and decreasing it during physical activity.  A complete  automatic control includes several risks for the patient.  High level of glucose (hyperglicemia) implies ketacidosis and low level (hypoglycemia) can be fatal leading to death. The AP controller must tolerate many unpredictable events
such as pump failures, sensor noise, meals and  physical activity.

AP Controller Falsification via SMT solver~\cite{shmarov_smt-based_2017} and robustness of STL~\cite{cameron_towards_2015} has been recently proposed.
In particular,  \cite{cameron_towards_2015} formulates a series of STL properties testing insulin-glucose regulatory system.  Here we show the advantages of using $\eSTL$ for this task. 
\paragraph{\bf PID Controller.}
Consider a system/process which takes as input a function $u(t)$ and produces as output a function $y(t)$. A PID controller is a simple closed-loop system aimed to maintain the output value $y(t)$ as close as possible to a set point $sp$.  It continuously monitors the error function, i.e., $e(t) = sp - y(t)$ and defines the input of the systems accordingly to $u(t) = K_p \cdot e(t) + K_i \cdot \int_{0}^t e(s)ds + K_d \cdot \frac{d}{dt}e(t)$.  The \emph{proportional} ($K_p$), \emph{integral} ($K_i$) and \emph{derivative} ($K_d$) parameters uniquely define the PID controller and have to be calibrated in order to achieve a proper behavior.  
\paragraph{\bf System.} PID controllers have been successfully used to control the automatic infusion of insulin in AP. In~\cite {shmarov_smt-based_2017}, for example, different PID have been synthesized to control the glucose level for the well studied Hovorka model~\cite{hovorka2004nonlinear}:
\begin{equation}
\label{eq:hovorka}
\frac{d}{dt}\vec G(t)= \vec F (\vec G(t), u(t),\vec \Theta),
\end{equation}
where the output $\vec G(t)$ represents the glucose concentration in blood and the input $u(t)$ is the infusion rate of bolus insulin which has to be controlled. The vector $\vec \Theta =(dg1,dg2,dg3,T_1,T_2)$  are the control parameters which define the quantity of carbohydrates $(dg1,dg2,dg3)$ assumed during the three daily meals and the inter-times between each of them $T_1$ and $T_2$. Clearly a  PID controller for Eq. \eqref{eq:hovorka} has to guarantee that under different values of the control parameters $\vec \Theta$ the glucose level remains in the \emph{safe region} $\vec G(t)\in [70,180]$.  In~\cite {shmarov_smt-based_2017}, four different PID controllers that satisfy the safe requirement, have been discovered by leveraging  SMT solver  under the assumption that the inter-times  $T_1$ and $T_2$ are both fixed to 300 minutes (5 hrs) and that $(dg1,dg2,dg3) \in (\mathcal{N}(40,10),\mathcal{N}(90,10),\mathcal{N}(60,10))$, which correspond to the average quantity of carbohydrates contained in breakfast, lunch and dinner\footnote{$\mathcal{N}(\mu ,\sigma^2)$ is the Gaussian distribution with mean $\mu$ and variance $\sigma ^2$.}. 
Here, we consider the PID controller $C_1$ which has been synthesized by fixing the glucose setting point $sp$ to $110\,mg/dl$ and maximizing the probability to remain in 
the safe region, provided a distribution of the control parameter $\vec \Theta$ as explained before.
We consider now some properties which can be useful to check expected or anomalous behaviors of an AP controller. 
\paragraph{\bf Hypoglycemia and Hyperglycemia.} 
Consider the following informal specifications: {\it never during the day the level of glucose goes under $70\, mg/dl$}, and {\it never during the day the level of glucose goes above $180\, mg/dl$}, which technically mean that the patient is never under Hypoglycemia or Hyperglycemia, respectively. These behaviours can be formalized with the two {\bf STL } formulas 
$\phi^{HO}_{STL} = \glob{[0,24h]} \vec G(t)\ge 70$ and $\psi^{HR}_{STL} =\glob{[0,24h]} \vec G(t) \le 180$. 
The problem of STL is that it does not distinguish if these two conditions are violated for a second, few minutes or even hours. It only says those events happen. 
Here we propose  stricter requirements described by the two following \textbf{$\eSTL$} formulas $\phi^{HO}_{SCL} =\kerSML{\mathtt{flat}}{[0,24h]}{0.95}{\vec G(t) \ge 70}$  for the Hypoglycemia regime, and $\phi^{HR}_{SCL} =\kerSML{\mathtt{flat}}{[0,24h]}{0.95}{\vec G(t) \le 180}$ for the Hyperglycemia regime. We are imposing not that globally in a day the hypoglycemia and the hyperglycemia event never occur, but that these conditions persist for at least 95\% of the day (i.e., 110 minutes). We will show above in a small test case how this requirement can be useful. 
\paragraph{\bf Prolongated Conditions.} As already mentioned in the motivating example, the most dangerous conditions arise when Hypoglycemia or Hyperglycemia last for a prolongated period of the day. In this context a typical condition is the  {\bf Prolongated Hyperglycemia} which happens if the total time under hyperglycemia (i.e., $\vec G(t) \ge 180$) exceed the 70\% of the day, or the  {\bf Prolongated Severe Hyperglycemia} when the level of glucose is above $300\,mg/dl$ for at least 3 hrs in a day. 
The importance of these two conditions has been explained in~\cite{sankaranarayanan_model-based_2017}, however the authors cannot formalized them in STL. On the contrary, $\eSTL$  is perfectly suited to describe these conditions as shown by the following two formulas:  $\phi^{PHR}_{SCL}  =\kerSML{\mathtt{flat}}{[0,24h]}{0.7}{\vec G(t) \ge 180}$ and $\phi^{PSHR}_{SCL}  = \kerSML{\mathtt{flat}}{[0,24h]}{0.125}{\vec G(t) \ge 300}$.
Here we use flat kernels to mean that the period of a day where the patient is under Hyperglycemia or Severe Hyperglycemia does not count to the evaluation of the boolean semantics. Clearly, an hyperglycemia regime in different times of the day can count differently. In order to capture this ``preference" we can use non-constant kernels.


\paragraph{\bf Inhomogeneous time conditions.} 
Consider the case of monitoring Hyperglycemia during the day. 
Even if avoiding that regime during the entire day is always a best practice, there may be periods of the day where avoiding it is more important than others. 
We imagine the case to avoid hyperglycemia with a particular focus on the period close to the first meal. We can express this requirement considering the following $\eSTL$ formula:  $ \phi^{PHR}_{SCL(Gauss)} = \kerSML{\mathtt{gauss}[0.03,0.1]}{[0,24h]}{0.07}{\vec G(t) \ge 180}$.
Thanks to an decreasing kernel, indeed, the same quantity of time under hyperglycemia which is close to zero counts more than the same quantity far from it. 

\paragraph{\bf Correctness of the insulin delivery.}
During the Hypoglycemia regime the insulin should not be provided.
The $\eSTL$ formula:   $ \glob{[0,24h]}(\kerSML{\mathtt{flat}}{[0,10 min]}{0.95}{\vec G(t)\le 70} \rightarrow \kerSML{\mathtt{flat}}{[0,10 min]}{0.90}{I(t)\le0})$
states that if during the next 10 minutes the patient is in Hypoglycemia for at least the 95\% of the time then the delivering insulin pump is shut off (i.e., $I(t)\le 0$) for at least the 90\% of the time. This is the ``cumulative" version of the STL property
$\glob{[0,24h]}(\vec G(t)\le 70 \rightarrow I(t)\le0)$ which says that in hypoglycemia regime no insulin should be delivered. 
During the Hyperglycemia regime the insulin should be provided as soon as possible.
The property $\eSTL$ formula: \ $ \glob{[0,24h]}({\vec G(t) \ge 300} \rightarrow \kerSML{\mathtt{exp}[-1]}{[0,10 min]}{0.9}{I(t)\ge k})$ says that if we are in severe Hyperglycemia regime (i.e., $\vec G(t) \ge 300 $) the delivered insulin should be higher than $k$ for at least the 90\% of the following 10 minutes.  We use a negative exponential kernel to express (at the robustness level) the preference of having a higher value of delivered insulin as soon as possible.

\begin{figure}[!t]
	\centering
	{\includegraphics[width=0.35\textwidth]{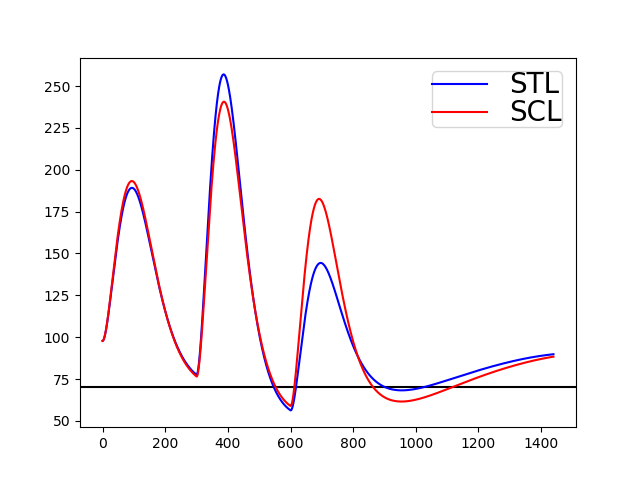}}
    \hspace{-0.5cm}
	{\includegraphics[width=0.35\textwidth]{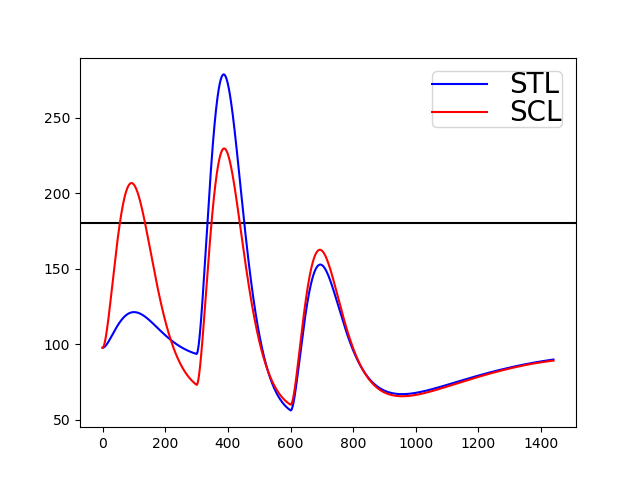}}
    \hspace{-0.5cm}
	{\includegraphics[width=0.35\textwidth]{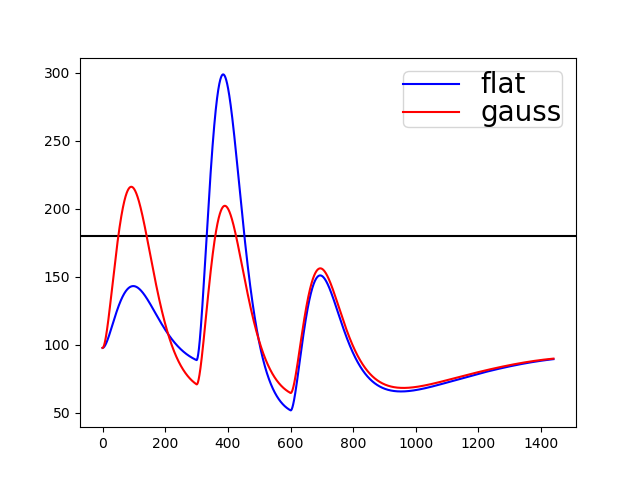}}
	\caption{ \textbf{(left),(middle)} The solution of the $\eSTL$ formula falsification (red line) maximize the time under Hypoglycemia (left) and Hyperglycemia (right), whereas the solution of the STL formula falsification (blue line) maximizes the displacement w.r.t the predicate thresholds.
	\textbf{(right)} Solution of the falsification for the $\eSTL$ properties $\phi^{PHR}_{SCL}$ (blue line) and $\phi^{PHR}_{SCL(Gauss)}$ (red line)  which implement \texttt{flat} and \texttt{gaussian} kernel, respectively.      
}
\vspace{-3ex}
\label{fig:battle}
\end{figure}

\paragraph{\bf Test Case: falsification.}
As a first example we show how $\eSTL$ logic can be effectively used for falsification. The AP control system has to guarantee that the level of glucose remains in a safe region, as explained before. The falsification approach consists in identifying the control parameters ($\vec \Theta^*$) which  force the system to violate the requirements, i.e., to escape from the safe region. The standard approach consists in minimizing the robustness of suited temporal logic formulas which express the aforementioned requirements, e.g., $\phi^{HR}_{SCL}, \phi^{HO}_{SCL}$. 
In this case the minimization of the STL robustness forces the identification of the control parameters which causes the generation of trajectories with a maximum displacement under the threshold $70$ or above $180$. 
To show differences among the STL and $\eSTL$ logics,  we consider the  PID $C_1$ + Hovorka model and perform a random sampling exploration among its input parameters. At each sampling we calculate the robustness of the STL formulas $\phi^{HO}_{STL}$ and the $\eSTL$ formula $\phi^{HO}_{SCL}$ and separately store the minimum robustness value. For this minimum value, we estimate the maximum displacement with respect to the hypoglycemia and hyperglycemia thresholds and the maximum time spent violating the hypoglycemia and hyperglycemia thresholds. Fig. \ref{fig:battle}(left, middle) shows the trajectory with minimum robustness. We can see that the trajectory which minimize the robustness of the STL formula has an higher value of the displacement from the hypoglycemia ($13$) and hyperglycemia ($98$) thresholds than $\eSTL$ trajectory (which are $11$ and $49$ respectively). On the contrary, the trajectory which minimizes the robustness of the $\eSTL$ formula remains under hypoglycemia (for $309$ min) and hyperglycemia (for $171$ min) longer than the STL trajectory ($189$ min and $118$ min, respectively).  These results show how the convolutional operator and its quantitative semantics can be useful in a falsification procedure.
This is particularly  evident in the Hyperglycemia case (Fig. \ref{fig:battle} (middle) ) where the falsification of the $\eSTL$ Hyperglycemia formula $\phi^{HR}_{SCL}$ shows two subintervals where the level of glucose is above the threshold.
In order to show the effect of non-homogeneous kernel, we perform the previous experiment, with the same setting, for  properties $\phi^{PHR}_{SCL}$  and $\phi^{PHR}_{SCL(Gauss)}$. 
From the results (Fig. \ref{fig:battle} (right)) is evident how the Gaussian kernel of property $\phi^{PHR}_{SCL(Gauss)}$ forces the glucose to be higher of the hyperglycemia threshold just before the first meal ($t \in [0,200]$) and ignores for example the last meal ($t \ge 600$).


\paragraph{\bf Test Case: noise robustness.}
Now we compare the sensitivity to noise of $\eSTL$ and STL formulae. We consider three degrees of hypoglycemia $h_k (t) =\{  G\le k\}$, where $k \in \{55,60,65,70\}$  and estimate the probability that the Hovorka model controlled by the usual PID $C_1$ (i.e., PID $C_1$ + Hovorka Model) satisfies the STL formulas $\phi_{STL}^k = \ev{[0,24h]}\,h_k$ and the $\eSTL$ formulas $\phi_{\eSTL}^k = \kerSML{\mathtt{flat}}{[0,24h]}{0.03}{h_k}$  under the usual distribution assumption for the control parameters $\vec \Theta$. The results are reported in column ``noise free" of Table \ref{tab:results_1}.
Afterwards, we consider a noisy outcome of the same model by adding a Gaussian noise, i.e., $\epsilon \in \mathcal{N}(0,5)$, to the generated glucose trajectory. We estimate the probability that this noisy system satisfies the STL and $\eSTL$ formulas above, see column ``with noise" of Table \ref{tab:results_1}. The noise correspond to the disturbance of the original signals which can occur, for example, during the measurement process.

As shown in Table \ref{tab:results_1}, the probability estimation of the STL formulas changes drastically with the addition of noise (the addition of noise forces all the trajectory to satisfy the STL formula). On the contrary, the $\eSTL$ formulas $\phi_{\eSTL}^k$ are more stable under noise and can be even used to approximate the probability of the STL formulas on the noise-free model. To better asses this, we checked how much the STL formula  $\phi_{STL}^k$ and the $\eSTL$ formula $\phi_{\eSTL}^k$, evaluated in the noisy model, agree with the STL formula $\phi_{STL}^k$ evaluated in the noise-free model, by computing their truth value on 2000 samples, each time choosing a random threshold $k\in [50,80]$. The score for STL is 56\%, while $\eSTL$ agrees on 78\% of the cases.

\begin{table}[!t]
	\centering
	\label{tab:results_1}
	\begin{tabular}{|c|c|c|c|c||c|c|c|c|c|}
		\hline
		& \multicolumn{4}{|c||}{noise free} & \multicolumn{4}{|c|}{with noise} \\
	    &$h_{55}$&$h_{60}$&$h_{65}$&$h_{70}$&$h_{55}$&$h_{60}$&$h_{65}$&$h_{70}$\\
		\hline
    	$\ev{[0,24h]} $&0.00&  0.19 & 0.81 & 1.00& 0.98 &1.00 &1.00 &1.00 \\
	\hline
	$\langle \mathtt{flat}_{[0, 24]}, 0.03  \rangle$	&0.00&0.00&  0.20& 0.91 & 0.00 &0.02 &0.77 &1.00 \\
	\hline
	\end{tabular}
\vspace{0.2cm}
\caption{Results of the falsification test case. The performance of STL and $\eSTL$ formulas verified on the PID $C_1$ + Hovorka model with noise and noise free are compared. The STL formula on the noisy model is uninformative.}
\vspace{-3ex}
\end{table}





\section{Conclusion}
\label{sec:conclusion}
We have introduced $\eSTL$, a novel specification language 
that employs signal processing operations to reason about temporal 
behavioural patterns.  The key idea is the definition of a family of 
modal operators which compute the convolution of a kernel 
with the signal and check the obtained value against a threshold.
Our case study on monitoring glucose level in artificial pancreas 
demonstrates how $\eSTL$ empowers the classical 
temporal logic operators (i.e., such as \emph{finally} and 
 \emph{globally}) with noise 
 filtering capabilities, and enable us to express temporal 
 properties with soft time bounds and with non symmetric 
 treatment of time instants in a unified way.  


The convolution operator of $\eSTL$ can be seen as a syntactic 
bridge between temporal logic and digital signal processing, trying 
to combine the advantages of both these two worlds.
This point of view 
can be explored further, bringing into the monitoring algorithms of 
$\eSTL$ tools from frequency analysis of signals.  
Future work includes the release of a Python library, and the 
design of efficient monitoring algorithms also for the quantitative 
semantics.  Finally, we also plan to develop online 
monitoring algorithms for real-time systems using hardware 
dedicated architecture such as field-programmable gate array (FPGA)
and  digital signal processor (DSP).

\bibliographystyle{splncs}
\bibliography{refs}
 

\end{document}